\documentclass[10pt]{iopart}
\usepackage{subfigure}
\usepackage{graphicx}
\usepackage{color}
\usepackage{bm}
\newlength{\figwidth}
\begin{document}
\title[Transport barrier for the radial diffusion due to the $E \times B$ drift]{Transport barrier for the radial diffusion due to the $E \times B$ drift motion of guiding centers in cylindrical confinement geometry}
\author{O. Izacard$^1$, N. Tronko$^1$, C. Chandre$^1$, G. Ciraolo$^2$, M. Vittot$^1$ \& Ph. Ghendrih$^3$}
\address{$^1$ Centre de Physique Th\'{e}orique - UMR 6207, CNRS/Aix-Marseille Universit\'{e}, Campus de Luminy - case 907, F-13288 Marseille cedex 09, France}
\address{$^2$ Laboratoire de M\'{e}canique, Mod\'{e}lisation \& Proc\'{e}d\'{e}s Propres - UMR 6181, Ecole Centrale Marseille, Technop\^{o}le de Ch\^{a}teau-Gombert, F-13451 Marseille cedex 20, France}
\address{$^3$ Institut de Recherche sur la Fusion Magn\'{e}tique - CEA/DSM Cadarache, F-13108 Saint-Paul-lez-Durance, France}
%\pacs{05.45.-a Nonlinear dynamics and nonlinear dynamical systems; 05.45.Gg Control of chaos, applications of chaos.}
\begin{abstract}
We consider the radial transport of test particles due to the ${\bf E}\times{\bf B}$ drift motion in the guiding center approximation. Using an explicit expression to modify the electrostatic potential, we show that it is possible to construct a transport barrier which suppresses radial transport. We propose an algorithm for the implementation of this local modification computed from an electrostatic potential known on a spatio-temporal grid. The number of particles which escape the inner region defined by the barrier measures the efficiency of the control. We show that the control is robust by showing a significant reduction of radial transport, when applied with a reduced number of probes aligned on a circle.
\end{abstract}

\section{Introduction}
Control of chaotic transport is an important topic in many areas of physics with considerable research and technical applications. Its aim is to reduce chaos when it is harmful and enhance it when it is beneficial. A lot of efforts have been devoted to reduce chaotic and turbulent transport in laboratory plasma physics. Two examples are afforded by the control of stochasticity of magnetic field lines and the transport generated by plasma instabilities. In magnetically confined fusion plasmas, like tokamaks, the electrostatic turbulence associated with the ${\bf E} \times {\bf B}$ drift motion, is one of the main sources for the loss of energy confinement ~\cite{Wootton_1990_PFB_2_2879,Scott_2003_PoP_10_963,Pettini_1988_PRA_38_344} and consequently constitutes a relevant obstacle to the attainment of plasma ignition. The development of control strategies able to induce a relevant change in transport properties by means of a small and localized perturbation constitutes a way to understand the complex behavior of laboratory plasmas.

A convenient way to control these systems is by using electromagnetic fields. Since the plasma dynamics is very sensitive to electrostatic fluctuations, a control method for the ${\bf E} \times {\bf B}$ drift motion of test particles has been proposed by a small and apt modification of the electrostatic potential~\cite{Ciraolo_2004_PRE_69_056213,Tronko_2009_JPA_42_085501}. This method has been tested on a Travelling-Wave Tube (TWT) for the reduction of chaotic transport~\cite{Chandre_2005_PRL_94_074101}. It is worth noticing that for these methods, the relevant modifications were applied in all the plasma region.
To step forward in the applicability of this control scheme to more complex plasma devices, we present a method and its practical algorithm to construct a transport barrier for the radial dynamics where the modification of the potential is only applied locally in a peripheral region of the plasma. Our strategy does not depend on the electrostatic potential at hand whether it is known analytically, numerically or experimentally. The geometry we are considering is a cylindrical spatial geometry which is shared by many linear devices like for example VINETA~\cite{Windisch_2006_PoP_13_122303}, MIRABELLE~\cite{Brochard_2005_PoP_12_062104} or CSDX~\cite{Burin_2005_PoP_12_052320}. Moreover, as in these devices the plasma temperature is not too high, plasma density and electrostatic potential are routinely measured by inserting probe arrays with large radial and poloidal extents. In that way it is possible to have a very accurate description of the space time behavior of the plasma~\cite{Grulke_2002_NJP_4_67,Windisch_2006_PS_T122_15,Windisch_2006_PoP_13_122303}. In addition it is worth noticing that it is also possible to act on the plasma using electrodes directly in contact with the plasma, producing for example a mode locking between the injected sinusoidal signal and the plasma dynamics~\cite{Klinger_1997_PPCF_39_B145,Brandt_2010_PPCF_52_055009}. Moreover, in the SOL region of tokamaks where the field lines are connected to the wall, a parallel current created by probes acts on perpendicular dynamics~\cite{Ghendrih_2003_NF_43_1013}.\\
Section 2 deals with the analytical solution of the potential to create a radial transport barrier for test particles. In Sec. 3, we describe the numerical algorithm to compute this modified potential from the electrostatic potential data. Finally, Section 4 is dedicated to the computation and the results of the control of test particle dynamics in a mock electrostatic potential.

\section{Method}
\subsection{Guiding-center dynamics}
The equations of motion of charged particles (of charge $e$ and mass $m$) in electromagnetic fields ($\mathbf{E}=-\mathbf{\nabla}V-\partial\mathbf{A}/\partial t$ and $\mathbf{B}=\mathbf{\nabla}\times\mathbf{A}$) can be obtained in a Hamiltonian framework \cite{Morrison_1982_AIP_88_13,Morrison_2005_PoP_12_058102} where the Hamiltonian is 
\begin{eqnarray*}
H(\mathbf{p},\mathbf{q},t) = \frac{\left( \mathbf{p} - e\mathbf{A}(\mathbf{q},t) \right)^{2}}{2m}+ e V(\mathbf{q},t),
\end{eqnarray*}
where ${\bf q}$ is the position of the particle and ${\bf p}$ is its canonically conjugate momentum. In strong magnetic fields, particles have a fast gyration motion around the guiding-center \cite{Northrop_1961_AP_15_79}. The relation between the position of the particle and its guiding center is:
\begin{eqnarray*}
\mathbf{q} = \mathbf{x} + {\bm{\rho}},
\end{eqnarray*}
where $\mathbf{x}$ is the position of the guiding-center and ${\bm \rho}$ is the Larmor vector. In what follows, we restrict our study to the case where the magnetic field is constant and uniform: ${\bf B}=B\hat{\bf z}$.
In this case, in order to reduce the system, we apply the generalized canonical transformation $(\bf{q},\bf{p})\mapsto(\bf{x},\bf{u})$ given by $\mathbf{x} = \mathbf{q} - \mathbf{\hat{z}} \times ({\bf p}-e{\bf A}) / eB,$ and $\mathbf{u} = ({\bf p} - e{\bf A}) / m$. The Hamiltonian becomes \cite{Littlejohn_1979_JMP_20_2445,Cary_2009_RMP_81_693}:
\begin{eqnarray*}
H =  \frac{m \mathbf{u}^{2}}{2} + e V\left( \mathbf{x}+\frac{m}{e B}\mathbf{\hat{z}} \times \mathbf{u} , t \right),
\end{eqnarray*}
and the Poisson bracket has pairs of conjugate coordinates: $(z,u_z)$, $(x,y)$ and $(u_x,u_y)$.
\begin{eqnarray*}
\{ F , G \} &=& -\frac{1}{e B} \hat{\bf z} \cdot \left( \frac{\partial F}{\partial \mathbf{x}} \times \frac{\partial G}{\partial \mathbf{x}} \right) + \frac{e B}{m^{2}} \hat{\bf z} \cdot \left( \frac{\partial F}{\partial \mathbf{u}} \times \frac{\partial G}{\partial \mathbf{u}} \right)\\
&& + \frac{1}{m} \left( \hat{\bf z} \cdot \frac{\partial F}{\partial \mathbf{x}} \hat{\bf z} \cdot \frac{\partial G}{\partial \mathbf{u}} - \hat{\bf z} \cdot \frac{\partial F}{\partial \mathbf{u}} \hat{\bf z} \cdot \frac{\partial G}{\partial \mathbf{x}} \right).
\label{EoM_Hamiltonian-structure}
\end{eqnarray*}
It is important to notice that the two coordinates ($x$ and $y$) of the guiding center in the plane transverse to the magnetic field are conjugate, as well as the two velocities ($u_x$ and $u_y$) of the particle in this transverse plane. Nothing is changed along the direction of the field lines. 
In order to reduce the dimensionality of our system, we approximate the potential by
\begin{eqnarray*}
V\left( \mathbf{x}+\frac{m}{eB}\mathbf{\hat{z}} \times \mathbf{u} , t \right) \approx V\left( \mathbf{x}, t \right), 
\end{eqnarray*}
which is a standard hypothesis for plasma devices with a strong magnetic field. In this way, we notice that the equations of motion for the position of the guiding center are decoupled from the ones of the velocity of the particle. The equations of motion are given by
\begin{eqnarray*}
\dot{x}&=&-\frac{1}{B}\frac{\partial V}{\partial y},\\
\dot{y}&=&+\frac{1}{B}\frac{\partial V}{\partial x},
\end{eqnarray*}
that is the velocity of the guiding centers is equal to the ${\bf E}\times {\bf B}$ drift velocity. 

We consider a cylindrical geometry as encountered in linear plasma devices. In this geometry, the polar coordinates are better suited. We perform the (non-canonical) change of coordinates $x=r\cos \theta$ and $y=r\sin\theta$. The Poisson bracket of the reduced system in the poloidal plane is changed into~\cite{Morrison_1998_RMP_70_467}
\begin{eqnarray*}
\{F,G\}=\frac{1}{reB}\left(\frac{\partial F}{\partial \theta}\frac{\partial G}{\partial r}-\frac{\partial F}{\partial r}\frac{\partial G}{\partial \theta} \right),
\end{eqnarray*}
with a reduced Hamiltonian $H=e \tilde{V}(r,\theta,t)$, where $\tilde{V}(r,\theta,t)=V(x,y,t)$. In what follows, we remove the tildas for simplicity of the notations. 
Hence, the equations of motion for the guiding centers are:
\begin{eqnarray}
&&\dot{r} = - \frac{1}{rB} \frac{\partial V}{\partial \theta},
\label{eq:EoM_gc_r}\\
&&\dot{\theta} = + \frac{1}{rB} \frac{\partial V}{\partial r}.
\label{eq:EoM_gc_theta}
\end{eqnarray}
The singularity at $r=0$ can be removed by reformulating the equations of motion in the canonically conjugate pair of variables $(\Psi,\theta)$ with $\Psi=r^2/2$. However the singularity is in general not a problem in the numerical integration of test particles since the measured fluctuating potentials vanish in the core region around $r=0$. The important region where confinement is crucial is the peripheral (edge) region.

\subsection{Control term for a transport barrier} 
We are looking at modifications of the Hamiltonian, i.e. of the electrostatic potential $V$, in order to create a radial transport barrier. We consider the following expression for the controlled potential $V_c$: 
\begin{equation}
V_c(r,\theta,t) = V\left( f(r,\theta,t) , \theta ,  t \right),
\label{eq:Vc_fct_V}
\end{equation}
where $f(r,\theta,t)$ is the control function that has to be determined. We define the map $T$ as:
\begin{equation}
\label{eq:T}
T: (r,\theta,t)\mapsto (f(r,\theta,t),\theta,t),
\end{equation}
with which the controlled potential is written as $V_c=V\circ T$. We impose that this new potential creates a transport barrier for the guiding centers located at $r=R(\theta,t)$. This is equivalent to imposing that $r=R(\theta,t)$ is invariant by the dynamics. More specifically, we have
\begin{eqnarray}
\dot{R} =  \partial_{t} R(\theta,t) + \dot{\theta} \partial_{\theta} R(\theta,t),
\label{rpoint}
\end{eqnarray}
where $\dot{R}$ and $\dot{\theta}$ are given by Eqs.~(\ref{eq:EoM_gc_r}) and (\ref{eq:EoM_gc_theta}) for the potential $V_c$ given by Eq.~(\ref{eq:Vc_fct_V}). Equation~(\ref{rpoint}) translates into
\begin{eqnarray*}
\frac{1}{RB}\frac{\partial V}{\partial r}\frac{\partial}{\partial \theta}\left[ f(R(\theta,t),\theta,t)\right]+\frac{\partial R}{\partial t}+\frac{1}{RB}\frac{\partial V}{\partial \theta}=0,
\end{eqnarray*}
where the derivatives of $V$ are taken at $r=f(R(\theta,t),\theta,t)$. This equation has two unknown functions $f$ and $R$ to be determined. We assume that the time-average of $V$ is zero for all $\theta$ at a given radius $r_0$. We obtain a solution for $f$ and $R$ assuming that
\begin{eqnarray}
f(R,\theta,t) = r_0,
\label{eq:f_fct_r0}
\end{eqnarray}
and 
\begin{eqnarray}
R(\theta,t) = \sqrt{ r_0^2 - \frac{2}{B} \Gamma \frac{\partial V}{\partial \theta}(r_0,\theta,t)},
\label{eq:R_fct_r0}
\end{eqnarray}
where $\Gamma$ is a linear operator acting on any function (with vanishing mean value) of time as the pseudo-inverse of the derivative with respect to time, e.g., acting on a function $\phi(t)$ as
\begin{eqnarray}
\displaystyle
\frac{\partial \left( \Gamma \phi(t) \right)}{\partial t}= \phi(t).\label{eq:Gamma}
\end{eqnarray}
There is still a lot of freedom for the choice of the control term, even if all these choices lead to the same equation for the transport barrier $R$ given by Eq.~(\ref{eq:R_fct_r0}). We solve Eq.~(\ref{eq:f_fct_r0}) using the following control terms:
\begin{eqnarray}
&& f_1(r,\theta,t) = r + \psi_1(\theta,t),
\label{eq:f1}\\
&& f_2(r,\theta,t) = r + e^{-(r-r_0)^2/2\sigma^2}\psi_2(\theta,t),
\label{eq:f2}
\end{eqnarray}
where the $\psi_i(\theta,t)$ are arbitrary functions of $\theta$ and $t$ to be determined by Eq.~(\ref{eq:f_fct_r0}). The first expression $f_1$ for the control term implies that the potential has to be modified on the entire phase space. However it is desirable to only apply the modifications on a particular region of the transverse plane. In order to localize radially the control required for creating the transport barrier, a first idea is to localize the potential modifications in a neighborhood of the barrier $R(\theta,t)$. The fact that this localization is varying in time makes this strategy challenging for an experimental realization. In order to address this issue we adopt a slightly different strategy: the control term $f_2$ leads to a modification of the potential only in the neighborhood of a circle of radius $r_0$ which is more adapted to an experimental application since the control is envisaged through a set of probes usually aligned along a circle, or a predefined fixed geometry (see below). A Gaussian function around the radius $r_0$ is used to describe this localization and $\sigma$ is related to the width at half maximum. Using Eqs.~(\ref{eq:Vc_fct_V}), (\ref{eq:f_fct_r0}) and (\ref{eq:R_fct_r0}), the controlled potentials associated with the control terms (\ref{eq:f1}) and (\ref{eq:f2}) are respectively:
\begin{eqnarray}
V_c(r,\theta,t) &=& V\left( r + r_0 - R , \theta , t \right),\label{eq:Vc1}\qquad\qquad\qquad\ \ \\
V_c(r,\theta,t) &=& V\left( r + (r_0 - R)e^{((R-r_0)^2-(r-r_0)^2)/(2\sigma^2)} , \theta , t \right),\label{eq:Vc2}
\end{eqnarray}
where $R(\theta,t)$ is defined by Eq.~(\ref{eq:R_fct_r0}).

\section{Numerical algorithm}
We investigate numerically the effect of the controlled potential $V_c=V\circ T$ on the dynamics of guiding centers. We have developed our algorithm such that it is applicable for any potential known on a spatio-temporal grid as obtained from numerical codes or as measured experimentally.

\subsection{Computation of the controlled electrostatic potential}
\begin{figure}
\centering
\includegraphics[width=\figwidth]{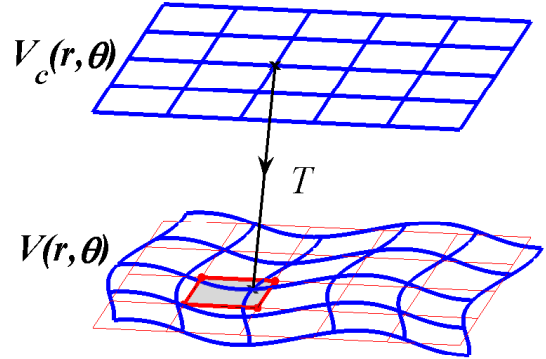}
\caption{The controlled potential $V_c$ on the initial grid is obtained by interpolating the uncontrolled potential $V$ (from the red grid) on  the deformed blue grid obtained from the map $T$ given by Eq.~(\ref{eq:T}).}
\label{fig:Grid-Modification_RT_3D}
\end{figure}

The measurement of the electrostatic potential $V$ is performed on a grid of $M_r\times M_{\theta} \times M_t$ points called the measurement grid. This is the input of our numerical algorithm. This measurement can be given by experimental measurements or numerical data. However the control term $f$ is computed on a refined spatial grid of $N_r\times N_{\theta} \times N_t$ points where $N_r > M_r$ and $N_{\theta} > M_{\theta}$ in order to minimize the error in the numerical scheme. The potential given by Eq.~(\ref{eq:Vc_fct_V}) is used in order to create a local barrier of radial transport. The controlled potential $V_c$ is computed on the measurement grid or on the refined one if the original grid is too coarse-grained (in order to avoid fake dissipation due to an imprecise way of computing Hamilton's equations). The output of this computation is the modifications of the electrostatic potential given on the measurement grid.\\
The derivatives with respect to periodic variables (e.g., $\theta$) as well as the $\Gamma$-operator given by Eq.~(\ref{eq:Gamma}) are computed using the Fast Fourier Transform and for non-periodic variables (e.g., $r$) with a fourth order finite-difference scheme.\\
As we can see from Eqs.~(\ref{eq:Vc1}) and~(\ref{eq:Vc2}), the controlled potential $V_c$ is simply obtained from the original one $V$ by a shift in the radial position $r$ (which depends on $\theta$ and $t$). This means that for determining $V_c$ on a given spatial grid, one has to recover by interpolation, the values of V on a deformed grid as schematically represented in Fig.~\ref{fig:Grid-Modification_RT_3D}. In this way, the method does not resort to an expansion of the potential and is more easily applicable to electric potentials obtained numerically or experimentally.

\subsection{Computation of test particles}
We compute and compare the dynamics of test particles given by the equations of motion~(\ref{eq:EoM_gc_r}) and~(\ref{eq:EoM_gc_theta}) obtained with the uncontrolled and controlled electrostatic potentials, $V$ and $V_c=V\circ T$ respectively. In order to integrate these equations of motion, we use a fourth order Runge-Kutta integrator. Even though this is not a symplectic scheme, we monitor the total energy $\mathcal{E} + V(x(t),y(t),t)$ where the dynamics of the additional variable $\mathcal{E}$ is defined by $\dot{\mathcal{E}} = -\partial V/\partial t$, so as to ensure that the numerical error is negligible. This also constitutes a way to check globally our numerical scheme. The dynamical equation for $\mathcal{E}$ is integrated with the same numerical scheme as Eqs.~(\ref{eq:EoM_gc_r}) and~(\ref{eq:EoM_gc_theta}). 

With a set of initial conditions inside the barrier $r=R(\theta, t)$, we compute the dynamics of test particles separately for the uncontrolled and for the controlled electrostatic potentials. If there is radial transport, the particles which are initially inside the barrier may leave the inner region. When the control creates a radial transport barrier, the test particles are expected to remain inside this barrier at all times.

\section{Numerical results}
\subsection{Computation with uncontrolled potential} 
\begin{figure}[!htp]
\centering
\includegraphics[width=\figwidth]{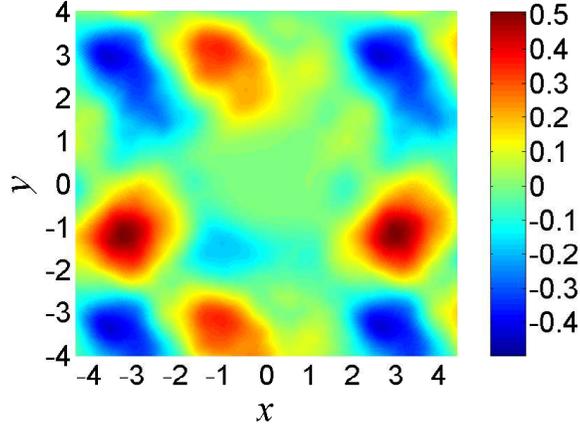}
\caption{Potential $V$ given by Eq.~(\ref{eq:Ciraolo2004PREPotential}) at $t=0$ for $a=0.4$. The initial condition of test particles positions are chosen randomly inside the barrier $R(\theta,0)$ which will be obtained for the controlled potential $V_c$ (see Fig.~\ref{fig:SPC-Controlled_f2}).}
\label{fig:H_RT}
\end{figure}

In order to validate our method with the computation of test particles, we use a toy analytic potential with a decay of the spatial modes in $k^{-3}$ consistent with experimental measurements~\cite{Truc_1984_PPCF_26_1045, Wootton_1990_PFB_2_2879}, which corresponds to the inertial range of the drift wave turbulence spectrum. We use only one frequency given that the frequency spectrum does not introduce changes in the algorithm as shown in Ref.~\cite{Ciraolo_2007_JNM_363_550}. This potential has been studied in Ref.~\cite{Pettini_1988_PRA_38_344} and is given by:
\begin{eqnarray}
\hspace{-2.2cm}\displaystyle
V(r,\theta,t) = a\sum_{\scriptsize\begin{array}{c} n,m=-N\\ n^2+m^2<N^2\\ n\neq 0, m\neq 0\end{array}}^{\scriptsize N} \frac{1}{ \left( n^2 + m^2 \right)^{3/2} } \sin\left( n r \cos\theta + m r \sin\theta + \varphi_{nm} - t \right),
\label{eq:Ciraolo2004PREPotential}
\end{eqnarray}
where $a$ is the amplitude of the potential and $\varphi_{nm}$ are random phases in order to model a turbulent potential with $N=25$. The amplitude $a$ is a parameter to modulate the importance of chaotic transport. Figure~\ref{fig:H_RT} is a representation of this potential at time $t=0$. We add a filter in a small circular region centered in $r=0$ to model the fact that the electrostatic fluctuations are weak at the center of the device.

In order to obtain an accurate numerical simulation, we choose the number of points $N_r = 256$ (in the radial position $r$), $N_{\theta} = 512$ (in the poloidal angle $\theta$) and $N_t = 64$ (in time) for the electrostatic potential grid. The measurement of the electrostatic potential is performed with $(M_r,M_{\theta})=(128,256)$ and $M_t=N_t$. The total energy conservation is accurate up to $10^{-4}$ up to a time of integration of $100$ periods. 
\begin{figure}[!htp]
\centering
\includegraphics[width=\figwidth]{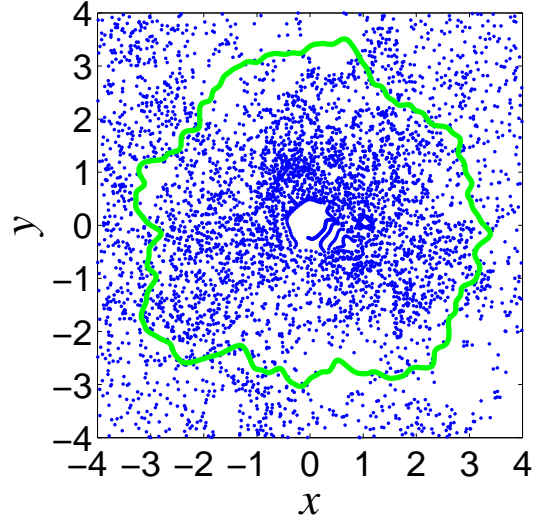}
\caption{Poincar\'{e} section of $100$ test particles for the potential $V$ given by Eq.~(\ref{eq:Ciraolo2004PREPotential}) for $a=0.4$, up to $t=100\times 2\pi$. The green curve is the position of the barrier $R(\theta,0)$ which will be obtained with the controlled potential $V_c$ (see Fig.~\ref{fig:SPC-Controlled_f2}).}
\label{fig:SPC-UnControlled}
\end{figure}
The initial positions of test particles are chosen near the center (randomly inside the position of the barrier $R(\theta,0)$ obtained later with the controlled electrostatic potential $V_c$, i.e. inside the interval $\left[ 0.45 ; 2.55 \right]$ where the minimum radius of the barrier will be approximately $2.6$). The Poincar\'{e} section (stroboscopic plot with the period of the potential) of test particles obtained with the uncontrolled electrostatic potential is shown in Fig.~\ref{fig:SPC-UnControlled}. The green curve defined by $r=R(\theta,t)$ at $t=0$ is added to show that guiding-centers can diffuse (without control) in all the available plane perpendicular to the magnetic field lines. More precisely, in the computation presented in Fig.~\ref{fig:SPC-UnControlled}, only $12$\% of the trajectories remain inside the barrier after $100$ periods of the potential.

\subsection{Computation with controlled potential}
\begin{figure}[!htp]
\centering
\includegraphics[width=\figwidth]{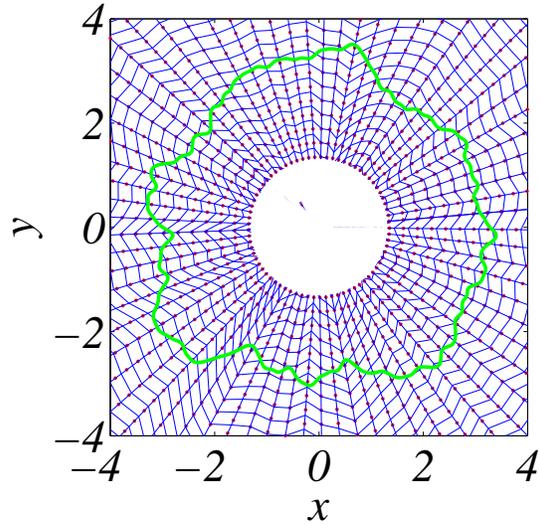}
\caption{Initial grid (red dots) where the uncontrolled potential $V$ is known. Modified grid (blue lines) defined by Eq.~(\ref{eq:T}) where $V$ is interpolated to obtain $V_c$ given by Eq.~(\ref{eq:Vc1}) on the initial grid (red dots). The green curve is the position of the barrier fluctuating near the fixed point $r_0=\pi$ at time $t=0$. We use the potential $V$ given by Eq.~(\ref{eq:Ciraolo2004PREPotential}) with $a=0.4$.}
\label{fig:Grid-Modification}
\end{figure}

\begin{figure}[!htp]
\centering
\includegraphics[width=\figwidth]{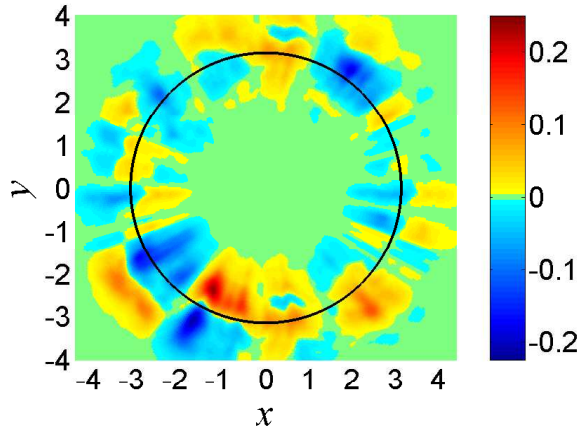}
\caption{Difference $V_c-V$ between $V_c$ given by Eq.~(\ref{eq:Vc2}) with $\sigma=0.6$ and $V$ given by Eq.~(\ref{eq:Ciraolo2004PREPotential}) at $t=0$ for $a=0.4$. This modification is centered around the black circle with radius $r_0=\pi$.}
\label{fig:H_RT_ctrl_f2}
\end{figure}

The control term and electrostatic potential are computed on the measurement grid $(M_r,M_{\theta})$. An additional step is to interpolate the controlled potential on the initial grid $(N_r,N_{\theta})$ using a spline interpolation. We apply our control algorithm in order to confine test particles inside the barrier. The controlled potential given by Eq.~(\ref{eq:Vc2}) is considered. It is computed numerically performing an interpolation method of $V$ on the deformed grid explicitly defined by $V_c=V\circ T$ following the procedure sketched in Fig.~\ref{fig:Grid-Modification_RT_3D}. For example, in Fig.~\ref{fig:Grid-Modification} we represent the deformation of the grid associated with the computation of the controlled potential given by Eq.~(\ref{eq:Vc1}). The difference $V_c-V$ where $V$ is given by Eq.~(\ref{eq:Ciraolo2004PREPotential}) and $V_c$ is given by Eq.~(\ref{eq:Vc2}) is shown in Fig.~\ref{fig:H_RT_ctrl_f2}. In this case, the maximum amplitude of the modification $V_c-V$ is less than $50$\% of the maximum amplitude of the potential $V$ for $a=0.4$.

\begin{figure}[!htp]
\centering
\includegraphics[width=\figwidth]{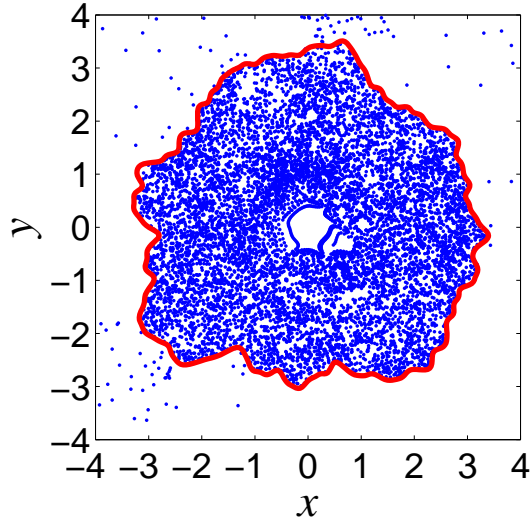}
\caption{Poincar\'{e} section of $100$ test particles for the potential $V_c$ given by Eq.~(\ref{eq:Vc2}) with $\sigma=0.6$ and $r_0=\pi$ where $V$ is given by Eq.~(\ref{eq:Ciraolo2004PREPotential}) for $a=0.4$ up to $t=100\times 2\pi$. The red curve is the position of the barrier $R(\theta,0)$.}
\label{fig:SPC-Controlled_f2}
\end{figure}

The efficiency of the control algorithm is estimated from the dynamics of test particles obtained by integrating Eqs.~(\ref{eq:EoM_gc_r}) and (\ref{eq:EoM_gc_theta}) with the controlled potential $V_c$. The Poincar\'{e} section of test particles obtained with the controlled potential $V_c$ given by Eq.~(\ref{eq:Vc2}) is shown in Fig.~\ref{fig:SPC-Controlled_f2}.

The position of the control barrier $R(\theta,2k\pi)=R(\theta,0)$ is represented by the solid red curve. We notice that most of the particles do not cross the transport barrier $r = R(\theta,t)$. A quantitative estimate of the efficiency of the control scheme is obtained by comparing the number of particles trapped inside the barrier for the controlled and uncontrolled potentials.
\begin{figure}[!htp]
\centering
\includegraphics[width=\figwidth]{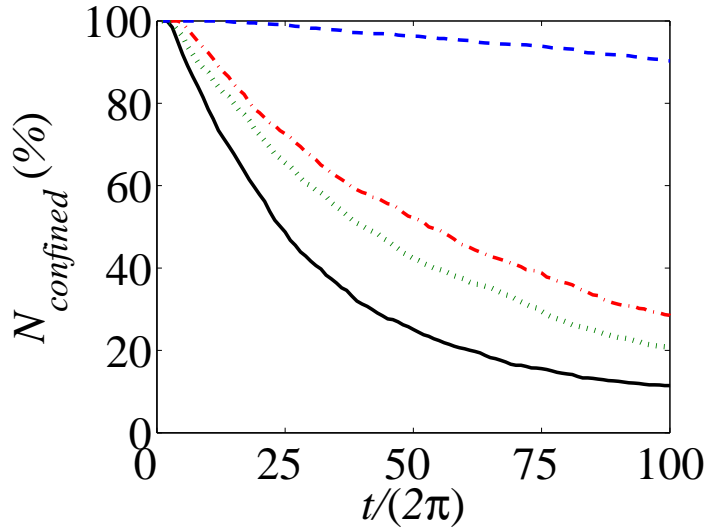}
\caption{Time evolution of the percentage of confined particles which stay inside the barrier i.e. with $(r,\theta)$ such that $r \leq \alpha R(\theta,t)$ where $\alpha=1.2$. The parameters of the potential are the same as in Fig.~\ref{fig:SPC-Controlled_f2}. This picture shows the results without (in solid black curve) or with the control (in dashed blue curve) computed from $2000$ particles. The control obtained with $32$ probes on a circle of radius $r=r_0$ is also shown (in dash-dotted red curve without measurement constraints or in dotted green curve with a measurement on $5 \times 128$ points).}
\label{fig:N_trapped}
\end{figure}
The time evolution of the percentage of particles $N_{confined}$ which stay inside the barrier is shown in Fig.~\ref{fig:N_trapped}. For $a=0.4$, up to $t=100\times 2\pi$, there is $12$\% of particles that remain confined within the prescribed region without the control (see Fig.~\ref{fig:SPC-UnControlled}) in comparison to $90$\% for the controlled potential (see Fig.~\ref{fig:SPC-Controlled_f2}). We notice that the control does not lead to $100$\% of trapped particles (the dashed blue curve in Fig.~\ref{fig:N_trapped}) as expected from Eq.~(\ref{eq:Vc2}) since the numerical algorithm (performed on a grid) induces an approximation of the exact control. This is also to be expected from the discrepancy between the measured potential and the actual one which prevents the computation of a very accurate controlled potential. However the fact that a significant number of test particles are trapped inside the transport barrier shows that the proposed control is robust in the numerical implementation. This robustness allows us to take into account experimental constraints. Here we have considered that the controlled potential is applied through a set of probes. Two constraints were considered: First, the geometry with which the probes are introduced should be as simple as possible. Second, the number of probes necessary for an effective implementation of the controlled potential should be relatively small and compatible with what is currently applied in linear plasma devices.\\
The first constraint is addressed by aligning the probes on a fixed circle of radius $r = r_0$ (see Fig.~\ref{fig:dH_probes}); the second one by using $M_{probes}=32$ probes and measuring its effects on the number of trapped particles. We assume that each probe creates an electrostatic potential with a Gaussian decrease of its influence in both the radial and poloidal direction. The potential created by the probe $i$ at the position $(r,\theta)=(r_0,\theta_i)$ is given by
\begin{eqnarray}
V_i(r,\theta,t) = A_i(t) e^{-(r-r_0)^2/2\sigma^2} e^{-(\theta-\theta_i)^2/2\sigma_{\theta}^2},
\end{eqnarray}
where $A_i(t)$ is the amplitude of the electric potential of the probe $i$ at time $t$ and $\sigma=0.6$ (respectively $\sigma_{\theta}=2\pi/M_{probes}$) is the spatial decrease of the electric potential in the radial (respectively poloidal) direction.\\
\begin{figure}[!htp]
\centering
\includegraphics[width=\figwidth]{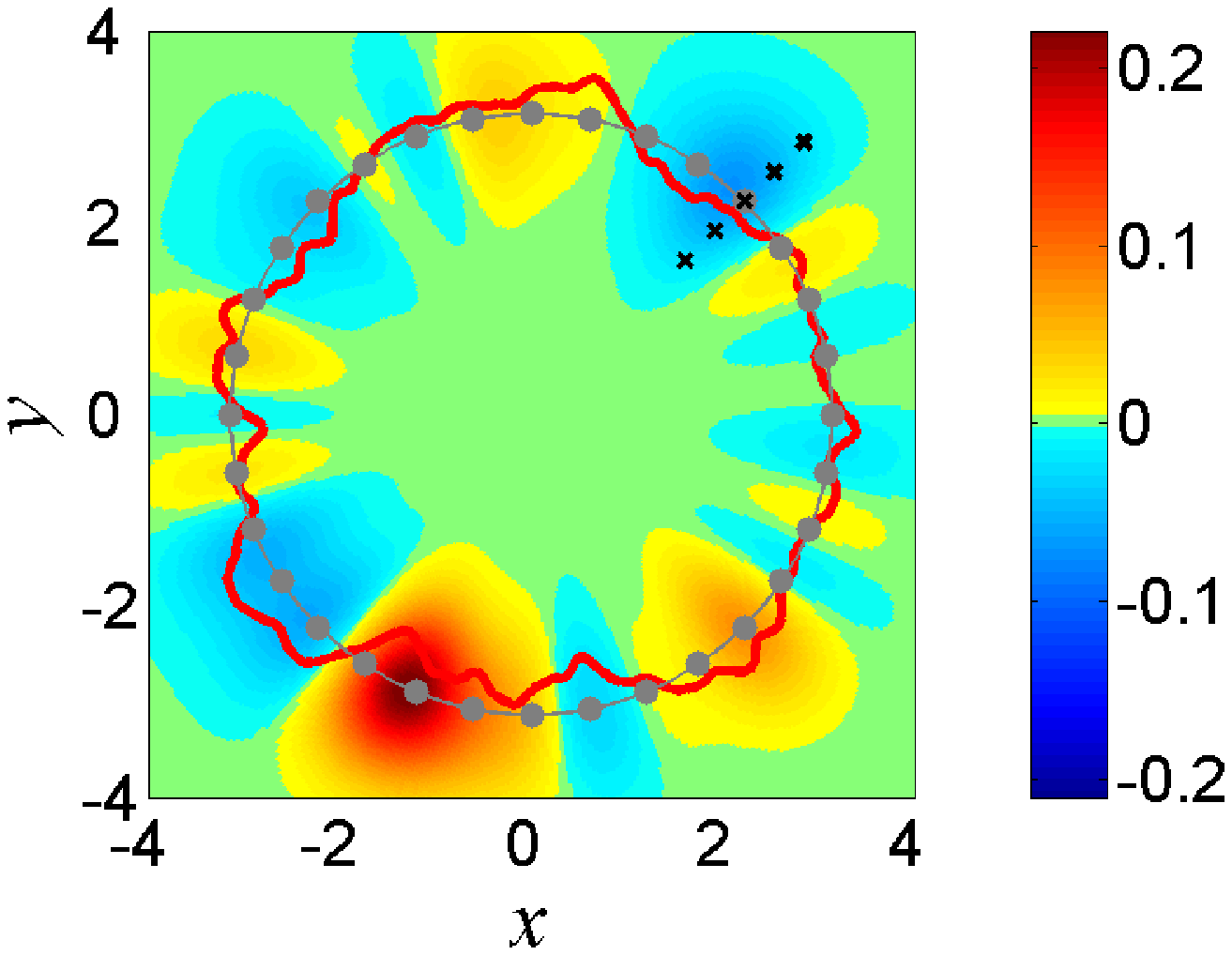}
\caption{The perturbations of the electric potential created by $32$ probes at $t=0$. The parameters of the potential are the same as in Fig.~\ref{fig:SPC-Controlled_f2}. The gray dots show the position of the probes and the solid red curve is the position of the barrier $R(\theta,0)$. The $5$ black crosses correspond to the position of one line of the measurement points which give the dotted green curve in Fig.~\ref{fig:N_trapped}.}
\label{fig:dH_probes}
\end{figure}
\begin{figure}[!htp]
\centering
\includegraphics[width=\figwidth]{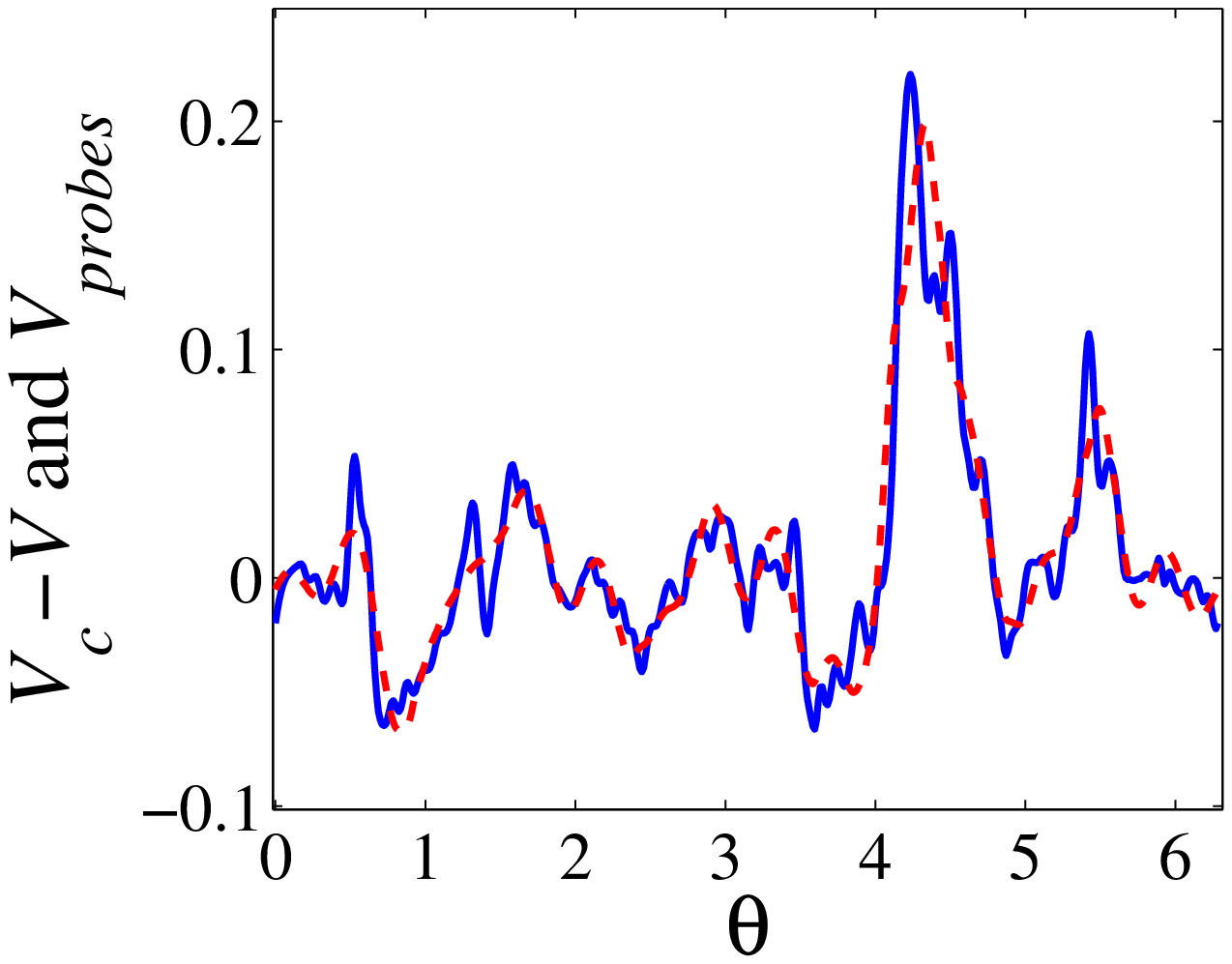}
\caption{Values of $V_c-V$ at $r=R(\theta,t)$ at time $t=0$ (in solid blue curve) and electric potential generated by $32$ probes (in dashed red curve). The parameters of the potential are the same as in Fig.~\ref{fig:SPC-Controlled_f2}.}
\label{fig:dH-on-R}
\end{figure}
If the potential introduced at $r = r_0$ is such that the implemented potential at $r = R(\theta,t)$ is as close as possible to the potential $V_c-V$ depicted in Fig.~\ref{fig:H_RT_ctrl_f2} (by adjusting the amplitudes $A_i(t)$) using the ordinary least square method, as shown in Fig.~\ref{fig:dH-on-R}, then our numerical results show that already with $32$ probes, the control strategy is effective in reducing radial transport as shown in Fig.~\ref{fig:N_trapped}. More quantitatively, we found that at $t = 100 \times 2\pi$, the number of trapped particles which is about $12$\% without the control is increased up to $29$\% with $32$ probes without measurement constraints and up to $21$\% with $32$ probes with a measurement on $5 \times 128$ grid points. In addition to robustness, an important advantage of the proposed strategy concerns the relatively small modifications of the potential. The mean average in angle $\theta$ and time $t$ (denoted $\langle\cdot \rangle$) of the potential modifications on the radial position $r=r_0$, i.e.\ $\langle \vert V(r_0,\theta,t)-V_c(r_0,\theta,t)\vert\rangle$, is about $30$\% of the mean average of the potential $\langle \vert V(r_0,\theta,t)\vert \rangle$ for $a=0.4$. This has to be compared with a simpler version of the control which consists in creating a potential barrier around $r=r_0$ such that $V_c(r=r_0,\theta,t)$ is small. In order to obtain the same quantity of trapped particles ($N_{confined} \sim 90$\% up to $t=100\times 2\pi$), it would be necessary to cancel $90$\% of the electrostatic potential which has to be compared with $30$\% of the modifications introduced by the proposed strategy with the same parameters as in Fig.~\ref{fig:SPC-Controlled_f2}.

\section*{Acknowledgments}
The authors would like to acknowledge the VINETA Nonlinear Plasma Dynamics Group of IPP-Greifswald for helpful discussions. We acknowledge financial support from the Agence Nationale de la Recherche (ANR EGYPT). This work was also supported by the European Community under the contract of Association between EURATOM, CEA, and the French Research Federation for fusion study. The views and opinions expressed herein do not necessarily reflect those of the European Commission.

\end{document}